# Instability of natural convection in a laterally heated cube with perfectly conducting horizontal boundaries

## Alexander Yu. Gelfgat

School of Mechanical Engineering, Faculty of Engineering, Tel-Aviv University, Ramat Aviv, Tel-Aviv, Israel, 69978, gelfgat@tau.ac.il

**Abstract**

Oscillatory instability of buoyancy convection in a laterally heated cube with perfectly thermally conducting horizontal boundaries is studied. The effect of the spanwise boundaries on the oscillatory instability onset is studied. The problem is treated by Krylov-subspace-iteration based Newton and Arnoldi methods. The Krylov basis vectors are calculated by a novel approach that involves the SIMPLE iteration and a projection onto a space of functions satisfying all linearized and homogeneous boundary conditions. The finite volume grid is gradually refined from $100^3$ to $256^3$ finite volumes. A self-sustaining oscillatory process responsible for the instability onset is revealed, visualized and explained.

## 1. Introduction

Buoyancy convection of air in a laterally heated square cavity is a widely recognized benchmark problem used for validation of numerical methods. It was proposed in [1] and since then is being mainly used for comparison of calculated steady flows. Later it was extended to compare calculated critical parameters of the primary steady – oscillatory transition. The reader is referred to book [2] for the details and the references. With the growth of computer power the two-dimensional formulation was replaced with the three-dimensional one, and the benchmark quality results for 3D steady flows are known and cross-verified [3-7]. Study of the primary instability of steady flows requires computation of the critical parameters via the comprehensive linear stability analysis, which involves direct computation of steady flows and eigenvalues of the linearized problem. This task has become affordable for 2D base flows (see [7-10] and references therein), but remains a challenge for fully three-dimensional problems.

Until now, the steady – oscillatory transition of buoyancy convection in a cube was studied by straight-forward time integration of the governing equations [11-14]. While this approach is the simplest, its disadvantages are well-known, and the results exhibit a considerable scatter [11]. This study presents the first attempt to study stability of buoyancy convection in a 3D laterally heated cube by direct Krylov-subspace-based Newton and Arnoldi solvers with the goal to obtain grid convergent stability results. We begin with a cubical box with perfectly thermally conducting horizontal boundaries, for which the critical Grashof number is relatively small [11,13]. The spanwise vertical boundaries are considered to be either perfectly thermally conducting of perfectly insulated, so that there are two similar, but different problems to study.

In the following, we formulate the problem and briefly describe the numerical method, as well as our method of visualization of three-dimensional divergence-free velocity field [15,16]. Evaluation of the Krylov vectors for BiCGstab(2), GMRES and Arnoldi Krylov-subspace-based iteration methods [17] is carried out using the novel approach. It is based on the SIMPLE iteration [18] for the velocity part and projection onto a suitable functional space for the temperature part of a Krylov vector. The SIMPLE procedure was already applied for this purpose in [19]. The projection of the temperature part of the Krylov vector onto the space of grid functions satisfying the boundary conditions is applied here for the first time.

In the "Results" section we start by reporting the convergence of the critical parameters varying the finite volume grid gradually from $100^3$ to $256^3$ nodes. Then we compare obtained 3D results with the known 2D ones and argue that transition to unsteadiness in the 2D and 3D cases is similar. Then, focusing on the two-dimensional case, we offer an explanation of the oscillatory self-sustaining process, which triggers the oscillatory instability. Considering then the three-dimensional case we identify a similar self-sustaining process, thus confirming the observed similarity. Using the above mentioned visualization, we describe also additional three-dimensional effects that appear in the fully 3D model.

In the course of results reporting and discussions, and comparison with fully non-linear time-dependent calculations of [11] we arrive to an assumption about a sub-critical character of the two 3D bifurcations computed, which is yet to be verified. We argue also that the previous conclusions about the Rayleigh-Bénard mechanism of these instabilities [10,13] are not completely correct. We show that the buoyancy effects play a crucial role in the instability, however the whole process differs from the classical Rayleigh-Bénard scenario.

## 2. Formulation of the problem

We consider natural convection of an incompressible fluid in a cubic cavity, whose opposite sidewalls are kept at constant and different temperatures $T_{hot}$ and $T_{cold}$, and the horizontal boundaries are perfectly thermally conducting, so that their temperatures attain a linear profile between $T_{cold}$ and $T_{hot}$. The flow is described by a set of Boussinesq equations that are rendered dimensionless taking the cube side length $H$ as a characteristic scale, and $H^2/\nu$, $\nu/H$, $\rho\,\nu^2/H^2$ as scales of the time $t$, the velocity $\boldsymbol{v} = (u, v, w)$ and the pressure $p$, respectively. Here $\nu$ is the fluid kinematic viscosity and $\rho$ is the density. The temperature is rescaled to a dimensionless function using the relation $T \rightarrow (T - T_{cold})/(T_{hot} - T_{cold})$. Additionally, the dimensionless time, velocity and pressure are scaled, respectively, by $Gr^{-1/2}$, $Gr^{1/2}$, and $Gr$, where $Gr = g\beta(T_{hot} - T_{cold})H^3/\nu^2$ is the Grashof number, $g$ is the gravity acceleration, and $\beta$ is the thermal expansion coefficient. The resulting system of momentum, energy and continuity equations is defined in a cube $0 \leq x, y, z \leq 1$ and reads

$$\frac{\partial T}{\partial t} + (\boldsymbol{v} \cdot \nabla)T = \frac{1}{PrGr^{1/2}}\Delta T \qquad (1)$$

$$\frac{\partial \boldsymbol{v}}{\partial t} + (\boldsymbol{v} \cdot \nabla)\boldsymbol{v} = -\nabla p + \frac{1}{Gr^{1/2}}\Delta\boldsymbol{v} + T\boldsymbol{e}_z \qquad (2)$$

$$\nabla \cdot \boldsymbol{v} = 0 \, . \qquad (3)$$

Here $Pr = \nu/\alpha$ is the Prandtl number, and $\alpha$ is the thermal diffusivity. All the boundaries are assumed to be no-slip. Two vertical boundaries at $x = 0,1$ are kept isothermal, so that

$$T(x = 0, y, z) = 1, \quad T(x = 1, y, z) = 0 \ . \tag{4}$$

The dimensionless linear temperature profile at the horizontal boundaries is

$$T(x, y, z = 0) = T(x, y, z = 1) = 1 - x \tag{5}$$

The two remaining boundaries, at $y = 0$ and 1, are called spanwise, and are assumed to be either perfectly thermally conducting or perfectly thermally insulated. Following the notations of [11], we denote a pair of perfectly conducting boundaries as CC, and a pair of perfectly insulating boundaries as AA. Thus, we arrive to two sets of thermal boundary conditions that are considered below

$$\text{CC} - \text{CC:} \qquad T(x, y = 0, z) = T(x, y = 1, z) = 1 - x \tag{6}$$

$$\text{CC} - \text{AA:} \qquad \left(\frac{\partial T}{\partial y}\right)_{y=0} = \left(\frac{\partial T}{\partial y}\right)_{y=1} = 0 \tag{7}$$

As is noted in [11,12], the problem has three symmetries: (i) reflection symmetry with respect to the midplane $y = 0.5$, $\{u, v, w, \theta\}(x, y, z) = \{u, -v, w, \theta\}(x, 1 - y, z)$, (ii) 2D rotational symmetry with respect to rotation in $180^\text{o}$ around the line $x=z=0.5$, $\{u, v, w, \theta\}(x, y, z) = -\{u, -v, w, \theta\}(1 - x, y, 1 - z)$ and (iii) 3D centro-symmetry $\{u, v, w, \theta\}(x, y, z) = -\{u, v, w, \theta\}(1 - x, 1 - y, 1 - z)$. Here, $\theta = T - (1 - x)$. These symmetries are characteristic for steady state flows, however, can be broken by instability, so that supercritical oscillatory flows can maintain only one of them or be fully non-symmetric. To compare results with the corresponding two-dimensional model of buoyancy convection in a square cavity, we call areas adjacent to the cube edges (0,y,0), (0,y,1), (1,y,0) and (1,y,1) as lower left, upper left, lower right, and upper right corners, respectively.

The primary goal of the $Gr^{1/2}$ scaling is to make the velocity values of the order of unity, which allows for more accurate calculations. The form of equations (1) and (2) shows also that $Gr^{1/2}$ yields an estimate of the Reynolds number, as is known for natural convection flows (see, e.g., [20]). As was shown in [21], the same scale is applied to the vertical velocity in the boundary layers developing near vertical heated walls. Also, as is argued in [13], the Brunt-Väisälä frequency is estimated also by $\sqrt{g\beta(T_{hot} - T_{cold})H}/H$, so that its dimensionless value obtained using the viscous time scale $H^2/\nu$ is $\sqrt{g\beta(T_{hot} - T_{cold})H^3/\nu^2} = Gr^{1/2}$. Thus, the dimensionless time is scaled additionally by the inverse Brunt-Väisälä frequency.

## 3. Numerical method

### 3.1. Calculation of the Krylov vectors

The governing equations were discretized by the finite volume method using the same schemes and the same staggered grid stretching as in [11]. The steady flows were calculated by the Newton method and the leading eigenvalue and eigenvector needed for the linear stability analysis were computed by the Chebyshev preconditioned Arnoldi method, using either the ARPACK package [22] or the EB13 module of the HSL library [23]. The Arnoldi method itself is a version of the Krylov subspace iteration, while corrections needed for the Newton method were calculated using either the restarted GMRES or BiCGstab(2) method, both based on the Krylov subspaces [17]. The Krylov basis vectors were calculated using the approach of [19] for the velocity part of the vector, and an additional treatment of the temperature part needed to preserve the temperature boundary conditions in every Krylov vector. The whole approach is briefly described below.

After the linearization, the continuity equation (3), the no-slip velocity boundary conditions, and the boundary conditions (4), (7) remain the same, while the boundary conditions (5) and (6) become homogeneous. The equations (1) and (2) linearized in the neighborhood of a steady flow denoted by $\boldsymbol{U}$, $P$, and $\Theta$ are

$$\frac{\partial T}{\partial t} + (\boldsymbol{U} \cdot \nabla)T + (\boldsymbol{v} \cdot \nabla)\Theta = \frac{1}{PrGr^{1/2}}\Delta T \tag{8}$$

$$\frac{\partial \boldsymbol{v}}{\partial t} + (\boldsymbol{U} \cdot \nabla)\boldsymbol{v} + (\boldsymbol{v} \cdot \nabla)\boldsymbol{U} = -\nabla p + \frac{1}{Gr^{1/2}}\Delta \boldsymbol{v} + T\boldsymbol{e}_z \tag{9}$$

Altogether, the linearized momentum equation, the continuity equation and the (linearized, if needed) homogeneous boundary conditions form the Jacobian operator $\mathfrak{J}$ that acts simultaneously on the velocity, pressure, and temperature fields. Then the Newton method aimed to computation of a steady state at a certain Reynolds number can be described as follows:

1. Choose an initial guess $(\boldsymbol{U}, P, \Theta)$;

2. Substitute $(\boldsymbol{U}, P, \Theta)$ into Eqs (1)-(4) and compute the residual vector $\boldsymbol{F}$; If $\|\boldsymbol{F}\| < \varepsilon$ exit.

3. Solve $\mathfrak{J} \begin{bmatrix} \boldsymbol{u} \\ p \\ \theta \end{bmatrix} = \boldsymbol{F}$

4. Make a new guess $\boldsymbol{U} \to \boldsymbol{U} - \boldsymbol{u}$, $P \to P - p$, $\Theta \to \Theta - \theta$ and go to the step 2.

The eigenvalue problem associated with the linear stability analysis reads

$$\mathfrak{J} \begin{bmatrix} \boldsymbol{u} \\ p \\ T \end{bmatrix} = \lambda \begin{bmatrix} \boldsymbol{u} \\ p \\ T \end{bmatrix}. \tag{10}$$

The steady flow $(\boldsymbol{U}, P, \Theta)$ is unstable if there exists at least one eigenvalue with a positive real part. In the following we call the eigenvalue with the maximal real part $\Lambda = \Lambda_r + i\Lambda_i$, as well as the associated eigenvector, "leading". The instability sets in at the critical value of Grashof number $Gr = Gr_{cr}$, at which $\Lambda$ crosses the imaginary axis, so that its real part turns from negative to positive. When the instability develops in a slightly supercritical regime, the most unstable disturbance oscillates with the critical circular frequency $\omega_{cr} \approx \Lambda_i$ and the exponentially growing amplitude is defined by the leading eigenvector.

Note that both the Newton method and the eigenvalue problem (10) treat the same Jacobian operator $\mathfrak{J}$. Assuming that the equations (1)-(4) and (6)-(7) are discretized by some numerical approach, the Jacobian operator reduces to a Jacobian matrix, which defines either the linear algebraic equation system of Step 3 of the Newton method, or the eigenvalue problem (10). These two problems are treated here by the Krylov-subspace iteration methods. Namely, the linear algebraic equations system is solved by either BiCGstab(2) or GMRES(n) [17], and the eigenvalue problem is treated by the Chebyshev preconditioned Arnoldi iteration aimed to finding the right-most eigenvalues [22,23].

Assume that the equations (1) – (7) are discretized by some numerical approach, and $\boldsymbol{v}^{(n)}, p^{(n)}, T^{(n)}$ form a current Krylov vector. The next vector, $\boldsymbol{v}^{(n+1)}, p^{(n+1)}, T^{(n+1)}$, must satisfy simultaneously

$$T^{(n+1)} = \frac{1}{PrGr^{1/2}} \Delta T^{(n)} - (\boldsymbol{U} \cdot \nabla) T^{(n)} - \left(\boldsymbol{v}^{(n)} \cdot \nabla\right)\Theta, \tag{11}$$

$$\boldsymbol{v}^{(n+1)} = -\nabla p^{(n+1)} + \frac{1}{Gr^{1/2}} \Delta \boldsymbol{v}^{(n)} - (\boldsymbol{U} \cdot \nabla) \boldsymbol{v}^{(n)} - \left(\boldsymbol{v}^{(n)} \cdot \nabla\right)\boldsymbol{U} + T^{(n)}\boldsymbol{e_z}, \tag{12}$$

$$\nabla \cdot \boldsymbol{v}^{(n+1)} = 0; \ + all \ the \ homogeneous \ boundary \ conditions \ . \tag{13}$$

If the expressions (13) are satisfied, the Krylov vectors belong to a subspace of divergence-free vectors satisfying all the homogeneous boundary conditions of the linearized problem. Contrarily, if, say, any of the constraints (13) are not satisfied, the Krylov iterations will seek the solution in a noticeably "wider" space, which in most cases leads to the loss of convergence. The most common way to overcome this difficulty is application of the Stokes preconditioning that can be computationally realized via carrying out the time steps of either full or linearized problems [24]. It was argued in [25] that the Stokes operator can be replaced by a more general one. This approach is effective for 2D problems, as well as for 3D stability problems with a periodic third direction, where the base flow remains two-dimensional. Applied to a fully three-dimensional problem, like the one considered here, this approach exhibits a slowed down convergence, especially when the grids are refined.

It was noted in [24] and later in [25] that a correct result $\boldsymbol{v}^{(n+1)}, p^{(n+1)}$ of the problem (12) and (13) can be interpreted as a projection of a vector $\boldsymbol{v}^{(n+1)}$ onto the above mentioned subspace of divergent free vectors satisfying all the linear and homogeneous boundary conditions. Similarly, the scalar field $T^{(n+1)}$ must be projected onto the subspace of scalar functions satisfying all the linearized homogeneous boundary conditions for temperature. We start a description of our computational procedure from numerical realization of the latter projection.

Assume that $\mathcal{L}$ is the Laplacian operator that acts on the temperature and includes all the homogeneous boundary conditions of the linearized problem. Following [26], we define this operator via Kronecker products of one-dimensional operators as follows. The second derivative operators that include the corresponding boundary conditions at the border nodes are $D_{xx}$, $D_{yy}$, and $D_{zz}$. They act on a row or a column of the grid function $f_{ijk} = f(x_i, y_j, z_k)$, where indices $i, j$, and $k$ vary between 1 and $N_x, N_y$ and $N_z$, respectively. Representing $D_{xx}, D_{yy}$ and $D_{zz}$ by matrices and following notations of the Kronecker (tensor) product, we represent the operator $\mathcal{L}$ as

$$\mathcal{L} = D_{xx} \otimes I_y \otimes I_z + I_x \otimes D_{yy} \otimes I_z + I_x \otimes I_y \otimes D_{zz} \tag{14}$$

where $I_x, I_y$ and $I_z$ are identity matrices of the order $N_x, N_y$ and $N_z$, respectively, and $\otimes$ denotes the tensor product. For the following we assume that the eigenvalue decompositions of matrices $D_{xx}, D_{yy}$ and $D_{zz}$ are known and are represented as

$$D_{xx} = E_x \Lambda_x E_x^{-1}, \quad D_{yy} = E_y \Lambda_y E_y^{-1}, \quad D_{zz} = E_z \Lambda_z E_z^{-1} \quad . \tag{15}$$

Here $E_x, E_y$ and $E_z$ are square matrices of the order $N_x, N_y$ and $N_z$, respectively, whose columns are eigenvectors of the matrices $D_{xx}, D_{yy}$ and $D_{zz}$. $\Lambda_x, \Lambda_y$ and $\Lambda_z$ are diagonal matrices having the eigenvalues of $D_{xx}, D_{yy}$ and $D_{zz}$ on their diagonals. According to [26], the operator $\mathcal{L}$ can be decomposed as

$$\mathcal{L} = \left(E_x \otimes E_y \otimes E_z\right)\widehat{\Lambda}\left(E_x^{-1} \otimes E_y^{-1} \otimes E_z^{-1}\right) \tag{16}$$

where

$$\widehat{\Lambda} = \left(\Lambda_x \otimes I_y \otimes I_z\right) + \left(I_x \otimes \Lambda_y \otimes I_z\right) + \left(I_x \otimes I_y \otimes \Lambda_y\right) \tag{17}$$

is a diagonal matrix of the order $N_x N_y N_z$ whose diagonal values are $\widehat{\Lambda}_{ijk} = \Lambda_{x,i} + \Lambda_{y,j} + \Lambda_{z,k}$. It is easy to see that the inverse operator $\mathcal{L}^{-1}$ is represented as [26]

$$\mathcal{L}^{-1} = \left(E_x \otimes E_y \otimes E_z\right)\widehat{\Lambda}^{-1}\left(E_x^{-1} \otimes E_y^{-1} \otimes E_z^{-1}\right) \tag{18}$$

which yields an analytical representation of the inverse Laplacian operator and can be used for a semi-implicit time integration [11,14].

For the purposes of the present calculations, we notice that the matrix $\left(E_x \otimes E_y \otimes E_z\right)$ contains the eigenvectors of the Laplacian operator that satisfy all the homogeneous boundary conditions and form a basis in the space of vectors of the $N_x N_y N_z$ length. Projection of the right hand side of Eq. (11) on this basis will yield the temperature part of the Krylov vector belonging to the needed space, i.e., satisfying all the homogeneous boundary conditions of the linearized problem.

The numerical realization of the above projection is complicated by the fact that the eigenvectors of a numerical approximation of the Laplacian operator are not necessarily orthogonal. Therefore, one needs either to apply an orthogonalization procedure, or to calculate the corresponding Gram matrix and its inverse. In the following, we choose the second option. Noticing that the Gram matrix $G$ can be defined via the Kronecker products as

$$G = G_x \otimes G_y \otimes G_z \quad , \tag{19}$$

where $G_x, G_y$, and $G_z$ are the Gram matrices of the one-dimensional bases $E_x, E_y$, and $E_z$, respectively. Thus, computations of the inverse of the Gram matrix is reduced to the three calculations of the inverse of the one-dimensional matrices

$$G^{-1} = G_z^{-1} \otimes G_y^{-1} \otimes G_x^{-1} , \tag{20}$$

that must be calculated only once and do not require a large computer memory to store them. Then the temperature part of the Krylov vector defined by Eq. (11) is replaced by

$$\widehat{T}^{(n+1)} = \left(E_x \otimes E_y \otimes E_z\right)\left(G_z^{-1} \otimes G_y^{-1} \otimes G_x^{-1}\right)T^{(n+1)}\left(E_x^T \otimes E_y^T \otimes E_z^T\right) \tag{21}$$

Where the superscript $T$ stays for the matrix transpose. The computational cost of calculation of the projection is 1.5 times larger than that of the TPF solver in [14,26] and can be estimated as $3N_x N_y N_z (N_x + N_y + N_z)$ multiplications.

For calculation of the velocity part of the Krylov vector, we apply the approach of [19], which is based on the SIMPLE [18] iteration. The projection of $\boldsymbol{v^{(n+1)}}$ on the subspace of divergence free vector functions satisfying all the boundary conditions is computed via the following algorithm.

Start with $\boldsymbol{v^{(n+1)}}$ from Eq. (10), and $p^{(n+1)} = 0$

Repeat until $\|\varphi\| < \varepsilon$

1. Solve $\Delta \varphi = \boldsymbol{\nabla} \cdot \boldsymbol{v}^{(n+1)}$; $\left[ \frac{\partial \varphi}{\partial n} \right]_\Gamma = 0$.

2. Correct $\boldsymbol{v^{(n+1)}}$: $\boldsymbol{v^{(n+1)}} \rightarrow \boldsymbol{v^{(n+1)}} - \boldsymbol{\nabla}\varphi$, $p^{(n+1)} = p^{(n+1)} + \varphi$

3. If boundary conditions for $\boldsymbol{v^{(n+1)}}$ are not satisfied, introduce them by changing the boundary values of $\boldsymbol{v^{(n+1)}}$ and go to stage 1.

Steps 1 and 2 of this algorithm are the Chorin projection [27] that yields a divergence free velocity field providing that the Poisson equation in Step 1 is solved precisely. At the boundaries, owing to the Newman boundary conditions for $\varphi$, this projection keeps the normal velocity component unchanged. However, it can alter the tangent component. Note also, that after calculation of $\boldsymbol{v^{(n+1)}}$ by Eq. (10), no boundary conditions are generally kept. At Step 3, the boundary conditions are restored by alteration of $\boldsymbol{v^{(n+1)}}$ boundary values. Obviously, this alters the divergence in next to the boundaries nodes, so that steps 1 and 2 should be repeated. It is easy to see that if the iterations converge, then the resulting fields $\boldsymbol{v^{(n+1)}}$ and $p^{(n+1)}$ are those required by Eqs. (12) and (13). Note that the above iterative procedure makes no assumptions regarding the problem or the numerical discretization scheme, except the assumption of incompressibility. It was successfully applied for the study of stability of a lid-driven flow in a cube [19], as well as for computation of divergence free projections for the visualization purposes [16]. In the present computations the convergence criterion was $|div\boldsymbol{v}| < 10^{-9}$ pointwise, and was reached in less than 10 iterations.

The above algorithm was applied in the following computations for generation of the Krylov basis for GMRES, BiCGstab(2) and Arnoldi methods. The GMRES method was restarted after each 100 Krylov vectors were computed. In the cases when the GMRES process saturated, its last approximation was supplied to the BiCGstab(2) as an initial guess, which finally yielded a converged solution.

Special attention was paid for an accurate solution of the Poisson equation at Step 1. An analytical TPF (Tensor Product Factorization algorithm) method [14], based on Eq. (18), was applied for that. Note that since the method is analytical, the solution $\varphi$ is obtained to within computer accuracy. Thus, in the computations below, after the Step 2 was completed, the maximal absolute values of the discretized divergence of $\boldsymbol{v}^{(n+1)}$ remained below $10^{-10}$. The TPF solver was slightly modified for computation of the projection of the temperature part of the Krylov vectors via Eq. (21). Note that this projection is also calculated to within the computer accuracy.

As it was emphasized in [19], the above way of calculating Krylov vectors allows one to apply the Arnoldi iteration in a mode that allows for calculation of the leading eigenvalue in a single run (see [22,23] for the details). This is an obvious advantage, which allowed us to complete the computations in all the cases reported below. It should be noticed, however, that convergence of the Arnoldi iteration in this mode can be very slow.

For all grids considered, the converged steady state is obtained after $6 - 8$ Newton iterations. At the same time, computation of the Newton correction at each iteration becomes noticeably more difficult with the grid refinement. Thus, for $100^3$ grid, it converges within 1000 BiCGstab(2) iterations. Starting from $150^3$ grid, the BiCGstab(2) iterations do not converge. To reach the convergence, we apply the GMRES with 100 Krylov vectors, restarting until it saturates. Then we use the saturated results as an initial guess for BiCGstab(2) and iterate until convergence. The GMRES method saturates after $2 - 7$ restarts for $150^3$, $5 - 20$ restarts for $200^3$, and $10 - 40$ restarts for $256^3$ grid. The BiCGstab(2) needs from 1000 to 5000 iterations to converge. After the steady state flow is computed, computation of the leading eigenvalue was performed using the module EB13 of the HSL library or the ARPACK library. One or two complex conjugated pairs of the leading eigenvalues were calculated. These calculations needed about 10,000 Arnoldi iterations for the $100^3$ grid, and about 80,000 iterations for the $256^3$ grid. Note that such a large amount of the Krylov subspace iterations becomes possible because of the described above fast calculation of a next Krylov vector. It should be mentioned also that the EB13 module, as well as the ARPACK library, does not allow one to alter the convergence criterion definition. Possibly, the total number of Arnoldi iterations can be reduced if it is relaxed accordingly to the computational needs. To ensure correctness of the eigenvalue/eigenvector calculations we monitored the relative residual $\|\boldsymbol{J}\boldsymbol{X} - \lambda\boldsymbol{X}\|/\|\boldsymbol{X}\|$, where $\lambda$ and $X = [u, p, T]^T$ are the eigenvalue and the eigenvector of Eq. (10). After the Arnoldi process was converged, the residuals always were below $10^{-10}$.

### 3.2. *Visualization of 3D flows*

To illustrate how different thermal boundary conditions affect the flow, we show the steady state isotherms at $Gr = 3.3 \times 10^6$, which is close to the critical value (see below) in Fig. 1. To represent three-dimensional velocity fields we implement the visualization method proposed in [15,16]. The 3D flow is visualized by divergence-free projections of the velocity field on three sets of coordinate planes, (*x,y*), (*y,z*), and (*x,z*). Namely, we compute three projections $\boldsymbol{v}_1$, $\boldsymbol{v}_2$, $\boldsymbol{v}_3$ of the velocity field $\boldsymbol{v}$ on subspaces formed by divergence free velocity fields having only two non-zero components in the coordinate directions. This allows us to visualize 3D flow by isosurfaces of the three scalar functions $\Psi_x$, $\Psi_y$, and $\Psi_z$, which are non-zero components of vector potentials of the above projections. The projection vectors are tangent to the isosurfaces so that visualization of a divergence-free three-dimensional flow can be done by three independent frames depicting the vector potentials and the velocity projections (see [15,16] for details).

Figures 2 and 3 illustrate the above visualization approach for steady flows calculated at slightly subcritical $Gr = 3.3 \times 10^6$ for CC – CC (Fig. 2) and CC – AA (Fig. 3) cases. As discussed in [15], where the visualization of similar flows for the perfectly thermally insulated horizontal boundaries was presented, the vector potential $\Psi_y$ represents the main convective circulation. The hot liquid ascends and descends along the isosurfaces of $\Psi_y$, which represent the main convective circulation. The three-dimensional addition to the main circulation, which takes place in the $(x, z)$ planes, is represented by additional vortical motion in the $(y, z)$ and $(x, y)$ planes which takes place along the isosurfaces of $\Psi_x$, and $\Psi_z$.

Looking at the patterns of the vector potentials (Figs. 2 and 3) it is easy to see similarities and differences of the velocity field in both cases considered. Thus we observe similarity of the patterns of $\Psi_z$ (cf. Figs. 2d and 3d). The patterns of $\Psi_y$ already exhibit some differences: in the CC – CC case (Fig. 2c), the circulations are located near the central part of the cube corners, while in the CC – AA case (Fig. 3c) they are shifted downwards near the hot wall $x = 0$, and upwards near the cold wall $x = 1$. The most striking differences are observed in the shapes of the isotherms of $\Psi_y$. While the outer envelopes of the main circulation (Figs. 2a and 3a) look similar, the flows inside them (Figs. 2b and 3b) are qualitatively different. The difference is obviously caused by different thermal boundary conditions (6) and (7). We discuss below how this difference reflects into the onset of instability in both flows.

## 4. Results

The grid stretching applied in all the calculations was the same as in [11]

$$\{x_i, y_j, z_k\} \rightarrow \{x_i, y_j, z_k\} - a \cdot sin[2\pi\{x_i, y_j, z_k\}] \,, \tag{22}$$

where the stretching parameter $a$ can be varied between 0 and 0.12. Most of the calculations are carried out for $a = 0.12$, which is the steepest stretching applied also in [11]. For the reasons explained below, some of the calculations were repeated for smoother stretchings.

It is well established that the primary transition in the 2D models takes place due to the Hopf bifurcation [28-32]. Also, in fully three-dimensional time-dependent computations the primary transition of flows is observed to be from a steady to a single-frequency oscillatory state, which also indicates on the Hopf bifurcation [11,13]. Thus, in the following calculations we monitor one or two pairs of complex conjugated dominant eigenvalues.

### 4.1. Critical Grashof numbers and oscillation frequencies

Critical Grashof numbers calculated on gradually refined grids are reported, together with the critical oscillation frequencies in Fig. 4. The critical frequencies (Fig. 4b) converge to within the fourth decimal place and are 0.2825 and 0.2637 for the CC – CC and CC – AA cases, respectively. The critical Grashof numbers, however, exhibit a qualitatively different convergence in the CC – CC and CC – AA cases. With the grid refinement, in the CC – AA case the critical numbers show a clear asymptotic convergence to the zero grid size, so that the Richardson extrapolation yields $Gr_{cr,CC-AA} = 3.395 \cdot 10^6$ (Fig. 4a). In the CC – CC case, and the stretching parameter $a = 0.12$, the critical Grashof numbers slowly grow with the grid refinement starting from the $100^3$ nodes grid and then for $150^3$, $200^3$, $210^3$, $220^3$, and $230^3$ grids (Fig. 4a), seemingly also showing the asymptotic behavior. However, for the finer grids with $240^3$, $250^3$ and $256^3$ nodes, the value of the critical Grashof number slightly decreases and remains unchanged at $Gr_{cr,CC-CC} = 3.3429 \cdot 10^6$ for the three finest grids applied. Due to computer limitations, we cannot refine the grid further. To get more insight in the convergence, we repeated the calculations for $a = 0, 0.05,$ and $0.1$. For all these stretchings the result for the $256^3$ grid arrived at the same value, while at the coarser grids, we did observe the asymptotically-like behavior (Fig. 4a). Note that the smallest and the largest critical values obtained for the grids finer than $200^3$ nodes are $3.429 \cdot 10^6$ and $3.441 \cdot 10^6$, so that the difference is below 0.5%. Since the critical frequencies and patterns of the leading eigenvectors remain identical for all the cases, this difference does not affect the further discussion.

In both CC – CC and CC – AA cases, the critical Grashof numbers, as well as the critical frequencies, are close to those corresponding to the instability of the two-dimensional laterally heated square cavity with perfectly conducting horizontal walls, which is $Gr_{cr,2D} = 2.969 \times 10^6$ and $f_{cr,2D} = 0.2535$ [9]. As will be shown below, patterns of the leading 3D and 2D perturbations are also similar. This similarity indicates on similar reasons that trigger the instability in 2D and 3D formulations, so that explanation of the flow destabilization should be similar in the two- and three-dimensional models.

Results of the direct numerical simulation for this case yield $Gr_{cr,DNS} \approx 3.3 \times 10^6$ for both cases, and $f_{cr,DNS} = 0.282$ for CC – CC and 0.264 for CC – AA cases, respectively [11]. The present linear stability analysis yields slightly larger critical Grashof numbers (see above). Taking into account that the time-dependent integration of [11] was done on the same grids and for the same discretization, these indicate on a subcritical perturbation. Furthermore, in both cases the dominant perturbations preserve all the three flow symmetries, while time-dependent computations [11] showed that in the oscillatory regimes all the symmetries are broken in the CC – AA case, and rotational symmetry and centro-symmetry are broken in the CC – CC case. These are another indication on the bifurcation subcriticality. It is worth mentioning that within the present approach we cannot study direction of bifurcation by means of bifurcation analysis as it was done in [28,29]. At the same time, repeating the 2D calculations using the global Galerkin method [28,29], we confirmed that the two-dimensional bifurcation is super critical. This shows that in spite of obvious similarities, transition to unsteadiness in the 2D and 3D cases can exhibit also a noticeable difference.

*4.2. Steady – oscillatory transition of the two-dimensional flow*

In this section we revisit the steady – oscillatory transition that takes place in the two-dimensional case. As noted above, the critical Grashof number and the pattern of leading perturbation are well established [9,30-32]. The absolute values of the 2D perturbations of the temperature and stream are shown by color plots in Fig. 5. The isotherms and the stream function of the base flow are depicted by black curves in the same figure. Note that the maximal values of temperature perturbation (Fig. 5a,b) are located in the lower left and upper right corners. Near these corners (Fig. 5b) we observe unstable temperature stratification, so that colder fluid is located above the isotherm 0.7 in the upper right corner. Similarly, hotter fluid is located below the isotherm 0.3 in the lower left corner. This observation lead to the assumption about the Rayleigh – Bénard instability mechanism that was first made in [13], and then repeated in many other studies, e.g., in [11,31].

At the same time maximal absolute values of the stream function disturbance (Fig. 5c,d) are located outside the unstable stratification area, at the border of the region where primary convective circulation splits into several secondary ones. The latter indicates on a possible hydrodynamic (i.e., not connected with the thermal convection) instability mechanism. Oscillations of the streamlines shown in Fig. 6 also exhibit most noticeable changes in the central part of the flow, where the temperature is stably stratified between the upper part of the isotherm $T = 0.3$ and the lower part of the isotherm $T = 0.7$, see Fig. 5b,d. The streamlines located near the boundaries, as well as in the unstably stratified region, are only slightly deformed over the period.

To gain more insight in the time evolution of the leading disturbance, we plot its history over the oscillation period in Figs. 7 and 8 (see also the corresponding animations). Note, that the disturbance preserves the rotational symmetry of the 2D flow, as it was previously reported in [32], where the most unstable perturbation mode was presented as Mode 1. We observe that amplitudes of both the stream function and the temperature perturbations start to grow in the lower left and upper right corners, however it is impossible to judge whether their growth is a result of the unstable stratification and the corresponding Rayleigh-Bénard instability mechanism.

A definite answer is obtained by the following computational experiment. Recalling that classical Raleigh-Bénard instability takes place owing to the velocity – temperature coupling terms $\theta \boldsymbol{e}_z$ in Eq. (9) and $w \, \partial T / \partial z$ in Eq. (8), we nullify these terms in the linearized eigenvalue problem and rerun the stability analysis. Assigning one or both of these terms to zero leads to increase of the critical Grashof number in more than an order of magnitude. This could already be an answer, but zeroing either of the terms $u\,\partial T/\partial x$, $W\,\partial\theta/\partial z$, or $U\,\partial\theta/\partial x$ in Eq. (8) also leads to a similar stabilization. At the same time, canceling of all the convective terms in the linearized momentum equation (9) only slightly alter the critical Grashof number and the imaginary part of the leading eigenvalue. Thus, zeroing the terms $(\boldsymbol{U}\cdot\nabla)\boldsymbol{v}$ and $(\boldsymbol{v}\cdot\nabla)\boldsymbol{U}$ in Eq. (9), and keeping all the convective terms in the energy Eq. (8), we obtain $\Lambda_r = 0$ at $Gr = 2.761\cdot10^6$ with $\Lambda_i = 2\pi\cdot 0.121$. We conclude that the observed instability is driven by the buoyancy force and convection of heat in both $x$ and $z$ directions, which appears to be more complicated than the classical Rayleigh – Bénard mechanism. At the same time, convection of momentum does not destabilize the flow.

To describe how the instability sets in, we return to Fig. 8 and the corresponding animation, showing evolution of the temperature perturbation over the oscillation period. We start from the frame denoted as $t = 3\tau/8$ ($\tau$ is the oscillations period), and assume that the temperature is slightly reduced in the lower left corner and/or slightly increases in the upper right corner. The latter corresponds to the maximum and minimum of the temperature disturbance in the frame. Obviously, this local decrease (increase) of the temperature slows down the local ascending (descending) motion along the hot (cold) walls, which happens owing to the buoyancy forcing term $\theta\boldsymbol{e}_z$. Consequently, motion along the horizontal borders towards the lower left and upper right corners also slows down, which leads to slight overheating of the fluid near the hot part of the lower horizontal boundary, and slight overcooling near the cold part of the upper boundary. These overheating and overcooling are seen in the appearance of additional maximum and minimum of disturbance in the frames denoted as $t = 4\tau/8$ and $t = 5\tau/8$. The temperature perturbation continues to be advected by the main circulation, so that a pair of the disturbance maximum and minimum observed in the frame $t = 3\tau/8$ dissipates, while the new pair form the pattern of frame $t = 7\tau/8$. This pattern is "opposite" to the one at $t = 3\tau/8$ which we started from, since it has a local maximum of the temperature perturbation in the lower left corner, and a local minimum in the upper right corner. It also is distanced at exactly a half a period from the frame $t = 3\tau/8$. Now the fluid is overheated in the lower left corner and overcooled in the upper right one, which increases the buoyancy force there and speeds up the local ascending / descending motion. As a result, the local motion along the horizontal boundaries toward the lower left and upper right corners also speeds up. The fluid near the hot / cold parts of the lower / upper horizontal boundaries moves faster and has less time to gain / lose heat, so that we observe formation of the disturbance minimum / maximum near the hot / cold parts of the lower / upper horizontal boundaries. Then the whole cycle repeats.

The above description is supported also by the patterns of the stream function disturbance oscillations shown in Fig. 7. Note that the perturbation forms a clockwise vortex around its negative values, and a counter-clockwise vortex around its positive values. Since the main convection loop is clockwise, as is shown in the upper frames of Figs. 7 and 8, negative perturbation values correspond to a local flow speed up, while positive values correspond to a local slowdown. Thus, starting from the time $t = 3\tau/8$ we predicted a slowdown in the lower left corner and upper right corners, and observe that a positive stream function disturbance appears and grows in these corners. These local positive perturbation maxima are advected by the main circulation and continue to grow until they reach their maximum at $t \approx \tau$ (coincides with $t \approx 0$). Also, as it was predicted above, the negative minima of the stream function disturbance appear in the lower left and upper right corners, starting to speed up the flow there. These minima are also advected by the main circulation reaching their least values at $t \approx \tau + 4\tau/8$.

### 4.3. Oscillatory instability of the three-dimensional flow

Absolute value of the temperature disturbance is shown in Fig. 9 for both CC – CC and CC – AA cases. Patterns of the absolute value are similar and resemble patterns of the perturbation amplitude reported in Fig. 3 of [11]. In both cases perturbation preserves all the three symmetries. Observation of the temperature perturbation patterns shown in Fig. 9 in the midplane $y = 0.5$ reveals a strong similarity with their two-dimensional counterpart shown in Fig. 5a,b. This supports the above assumption about the similarity of 2D and 3D transitions, which was made on the basis of closeness of the 2D and 3D critical values. Thus, the self-sustained oscillatory mechanism in the 3D cases is expected to be similar to the 2D one, which is illustrated and discussed below.

Figures 10 and 11 show time history of the temperature perturbation in the CC – CC and CC – AA cases, respectively. Similarly to the two-dimensional case, we observe perturbation maximum and minimum appearing in the lower left and upper right corners and advected by the main circulation (cf. Figs. 10 and 11 with Fig. 8 and the corresponding animations), thus resembling observation of the 2D model. Oscillations of the disturbance of the velocity potential $\Psi_y$, which is analogous to the 2D stream function [15,16], are shown in Fig. 12 for the CC – CC case. For the CC – AA case these oscillations look similar. Arrow plots in two frames show the direction of the disturbed flow part. Comparing Figs. 7 and 12, we again observe clear similarity between instability onset in the 2D and 3D models.

In spite of the clear similarity between the instability onset in the 2D and 3D cases, there are still some three-dimensional additions to the 3D oscillatory flow state that are shown in Figs. 13 and 14. The patterns in these figures are antisymmetric with respect to the $y = 0.5$ midplane because they contain projections of the $y$- velocity component, which is also antisymmetric. Therefore, all the flow symmetries are preserved. Note also, that vector plots added to several frames of Figs. 13 and 14, as in Fig. 11, indicate on the directions of perturbed part of the flow only.

Figure 13 shows that disturbances of ascending / descending motion along the heated and cooled borders, which are maximal at the midplane $y = 0.5$, create vortices not only in the $(x, z)$, but also in the $(y, z)$, planes. The latter are located in the boundary layers adjacent to the heated and cooled walls (Fig. 13). Note that intensities of the potentials $\Psi_y$ and $\Psi_x$ are comparable, so that the flow being pushed up or down along the centerline returns back in all possible directions. This is clearly different from the 2D case where the velocity vector cannot leave the $(x, z)$ plane.

Figure 14 shows disturbances of 3D flow motion in the $(x, y)$ planes. The corresponding vector potential $\Psi_z$ is noticeably weaker than $\Psi_y$ and $\Psi_x$ so that this part of the motion can be considered as secondary. We observe here that the vortical motion in the $z = const$ planes also speeds up or slows down the flow in the midplane $y = 0.5$. However, contrary to two former figures, the flow is affected along the horizontal boundaries in the upper left and lower right corners, which are opposite to the corners where we observed the instability onset. This can be interpreted as a reaction to the perturbed flow advected towards these corners.

## 5. Conclusions

In this study, we report the values of the critical Grashof number and critical frequency calculated for two different configurations of the three-dimensional buoyancy convection flow in a laterally heated cubical cavity with perfectly conducting horizontal boundaries, and either conducting or insulating spanwise boundaries. These values are calculated via comprehensive linear stability analysis that included the direct computation of steady state flows, followed by computation of the leading eigenvalue/eigenvector of the linearized stability problem.

The Newton iteration based steady state solver involved Krylov-subspace iteration methods, BiCGstab(2) and restarted GMRES, for solution of the linear algebraic equations required at each iteration. For all the Krylov-subspace iterations methods applied, we used the recently proposed approach of [19] for calculation of divergence-free Krylov vectors that satisfy all the linear (or linearized) homogeneous boundary conditions. The Krylov vectors were calculated using the SIMPLE-like procedure [18]. In this study we complemented the above procedure by calculation of the temperature part of the Krylov vectors, which also satisfies the linearized homogeneous boundary conditions. For this purpose, we built a projection operator that projects arbitrary vectors on the subspace of functions satisfying the boundary conditions needed. The whole approach for calculation of the Krylov vectors does not depend on a problem or on a numerical discretization, so that its applications can be rather wide. Furthermore, the Arnoldi iteration can be carried out in the mode that allows for the direct calculation of the dominant eigenvalue and eigenvector. The latter allowed us to complete the 3D computations on the series of gradually refined grids consisting of $100^3$ to $256^3$ finite volumes.

Another result of this study is the description of the self-sustained physical process that leads to the oscillatory instability of the flow. First, we have shown that in the 2D and 3D models the instability sets in a similar way. Then, analyzing the contribution of different terms of the governing equations in the onset of two-dimensional instability, we concluded that only buoyancy force and advection of heat play a significant role there. A closer look at the time evolution of the temperature disturbance allowed us to offer an explanation of the self-sustained oscillatory process. Furthermore, we showed that a similar mechanism is observed also in the two three-dimensional configurations considered. We discussed also secondary three-dimensional effects of the instability onset that necessarily appear in 3D problems.

**ACKNOWLEDGMENTS**

This research was supported by Israel Science Foundation (ISF) grant No  415/18 and was enabled in part by support provided by WestGrid ([www.westgrid.ca](www.westgrid.ca)) and Compute Canada (www.computecanada.ca).

**Figure captions**

Figure 1. Isotherms $T = 0.25, 0.5$ and $0.75$ for the CC – CC (a) and CC – AA cases at $Pr = 0.71$ and $Gr = 3.3 \times 10^6$.

Figure 2. Visualization of 3D velocity fields corresponding to slightly subcritical steady states at $Pr = 0.71$, $Gr = 3.3 \times 10^6$ by divergence free projections of velocity fields on the coordinate planes. CC – CC case. The projected velocities fields are depicted by vectors. Isosurfaces of the velocity potentials, to which the projected velocities fields are tangent, are shown by colors. The minimal and maximal values of the potentials are $\pm 0.0148$, $(-0.0572, 0.000487)$ and $\pm 0.0144$ for $\Psi^{(x)}$, $\Psi^{(y)}$, and $\Psi^{(z)}$, respectively. The isosurfaces are plotted for the levels $\pm 0.01$ for $\Psi^{(x)}$ (c), $\pm 0.006$ for $\Psi^{(z)}$ (d), and $-0.0185$ (a) and $-0.023$ (b) for $\Psi^{(y)}$.

Figure 3. Visualization of 3D velocity fields corresponding to slightly subcritical steady states at $Pr = 0.71$, $Gr = 3.3 \times 10^6$ by divergence free projections of velocity fields on the coordinate planes. CC – AA case. The projected velocities fields are depicted by vectors. Isosurfaces of the velocity potentials, to which the projected velocities fields are tangent, are shown by colors. The minimal and maximal values of the potentials are $\pm 0.00850$, $(-0.0265, 0.000182)$ and $\pm 0.00724$ for $\Psi^{(x)}$, $\Psi^{(y)}$, and $\Psi^{(z)}$, respectively. The isosurfaces are plotted for the levels $\pm 0.004$ for $\Psi^{(x)}$ (c), $\pm 0.003$ for $\Psi^{(z)}$ (d), and $-0.018$ (a) and $-0.023, -0.0205$ (b) for $\Psi^{(y)}$.

Figure 4. Convergence of the critical Grashof numbers and critical frequencies with the grid refinement. $N$ denotes the number of grid points in one spatial direction.

Figure 5. Color plots: amplitude of the temperature (a), (b) and stream function (c), (d) perturbations. Black lines: streamlines (a), (c) and isotherms (b), (d) of the base flow at the critical Grashof number. $Pr = 0.71, Gr_{cr} = 2.9 \times 10^6$. Minimal and maximal values of the stream function in the frames (a) and (c) are $-36.86$ and $0$.

Figure 6. Snapshots of the streamlines of slightly supercritical oscillatory 2D flow obtained as a superposition of the base state with the leading disturbance. The streamlines are equally spaced between the values $-0.026$ and $0$. Axes like in Fig. 4. Animation 1.

Figure 7. Snapshots of 2D dominant perturbation of the stream function shown on the streamlines (upper row) and isotherms (lower row) of the base 2D state at the critical point. Stream function is equally spaced between the values $-35$ and $0$, and the temperature between $0$ and $1$. Axes like in Fig. 4. Animation 2.

Figure 8. Snapshots of 2D dominant perturbation of the temperature shown on the streamlines (upper row) and isotherms (lower row) of the base 2D state at the critical point. Stream function is equally spaced between the values $-35$ and $0$, and the temperature between $0$ and $1$. Axes like in Fig. 4. Animation 3.

Figure 9. Amplitude of the temperature perturbations with isotherms of the base flow at the critical Grashof number. $Pr = 0.71, Gr_{cr} = 3.4 \times 10^6$. (a) CC – CC case, (b) CC – AA case .

Figure 10. Snapshots of 3D perturbation of the temperature with isotherms of the base flow in the midplane. CC – CC case. Axes like in Fig. 8.  Animation 4.

Figure 11. Snapshots of 3D perturbation of the temperature with isotherms of the base flow in the midplane. CC – AA case. Axes like in Fig. 8.  Animation 5.

Figure 12. Snapshots of 3D perturbation of the vector potential $\Psi_y$ with isotherms of the base flow in the midplane. $max|\Psi_z| = 2.4 \cdot 10^{-4}$. Levels shown are $\pm 10^{-4}$.  Vector plots indicate directions of perturbed part of the flow only.   CC – CC case. Animations 6 and 7.

Figure 13. Snapshots of 3D perturbation of the vector potential $\Psi_x$ with isotherms of the base flow in the midplane. $max|\Psi_x| = 2.5 \cdot 10^{-4}$. Levels shown are $\pm 10^{-4}$.  Vector plots indicate directions of perturbed part of the flow only.   CC – CC case. Animations 6 and 7.

Figure 14. Snapshots of 3D perturbation of the vector potential $\Psi_z$ with isotherms of the base flow in the midplane. Vector plots indicate directions of perturbed part of the flow only.  CC – CC case. $max|\Psi_z| = 7.5 \cdot 10^{-5}$. Levels shown are $\pm 2 \cdot 10^{-5}$.    Animations 6 and 7.

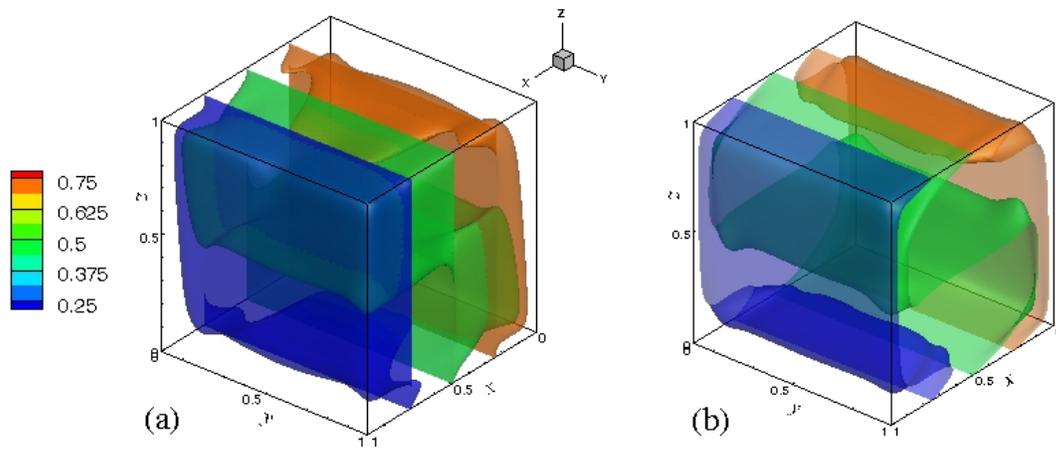

Figure 1. Isotherms $T = 0.25, 0.5$ and $0.75$ for the CC – CC (a) and CC – AA cases at $Pr = 0.71$ and $Gr = 3.3 \times 10^6$.

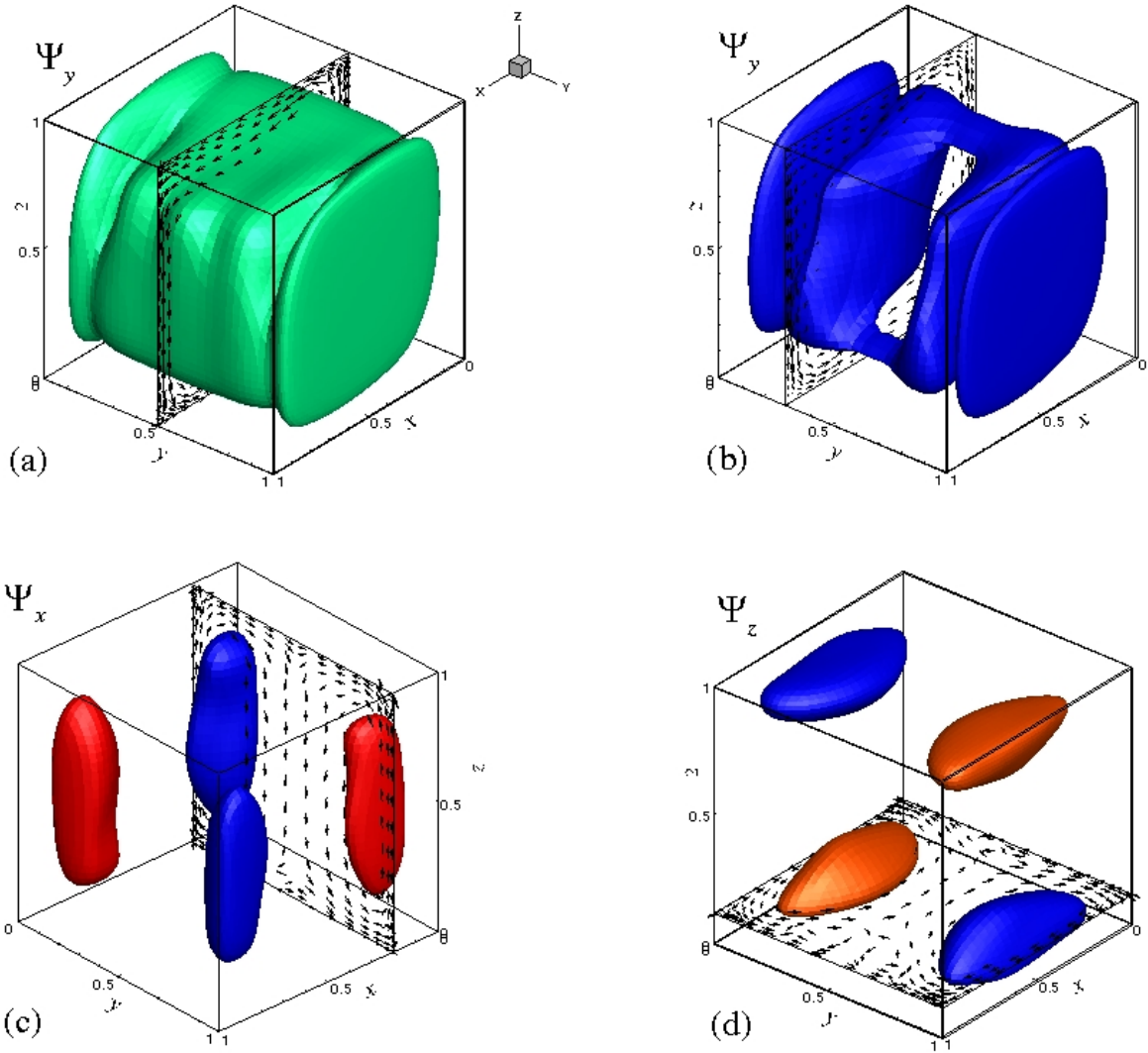

Figure 2. Visualization of 3D flow corresponding to slightly subcritical steady states at $Pr = 0.71$, $Gr = 3.3 \times 10^6$ by divergence free projections of velocity field on the coordinate planes. CC – CC case. The projected velocities fields are depicted by vectors. Isosurfaces of the velocity potentials, to which the projected velocities fields are tangent, are shown by colors. The minimal and maximal values of the potentials are $\pm 0.0148$, $(-0.0572, 0.000487)$ and $\pm 0.0144$ for $\Psi^{(x)}$, $\Psi^{(y)}$, and $\Psi^{(z)}$, respectively. The isosurfaces are plotted for the levels $\pm 0.01$ for $\Psi^{(x)}$ (c), $\pm 0.006$ for $\Psi^{(z)}$ (d), and $-0.0185$ (a) and $-0.023$ (b) for $\Psi^{(y)}$.

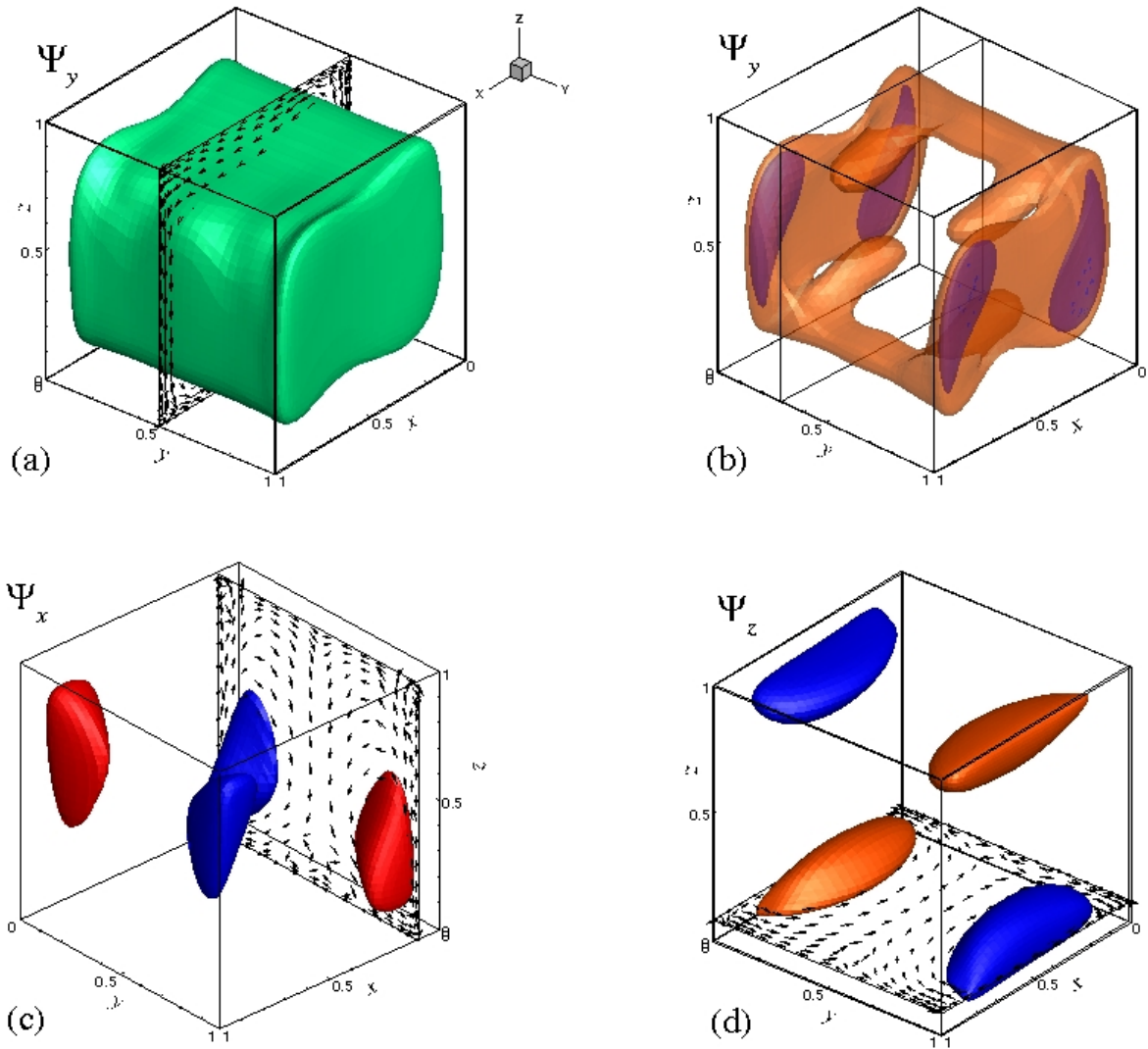

Figure 3. Visualization of 3D flows corresponding to slightly subcritical steady states at $Pr = 0.71$, $Gr = 3.3 \times 10^6$ by divergence free projections of velocity field on the coordinate planes. CC – AA case. The projected velocities fields are depicted by vectors. Isosurfaces of the velocity potentials, to which the projected velocities fields are tangent, are shown by colors. The minimal and maximal values of the potentials are $\pm 0.00850$, $(-0.0265, 0.000182)$ and $\pm 0.00724$ for $\Psi^{(x)}$, $\Psi^{(y)}$, and $\Psi^{(z)}$, respectively. The isosurfaces are plotted for the levels $\pm 0.004$ for $\Psi^{(x)}$ (c), $\pm 0.003$ for $\Psi^{(z)}$ (d), and $-0.018$ (a) and $-0.023$, $-0.0205$ (b) for $\Psi^{(y)}$.

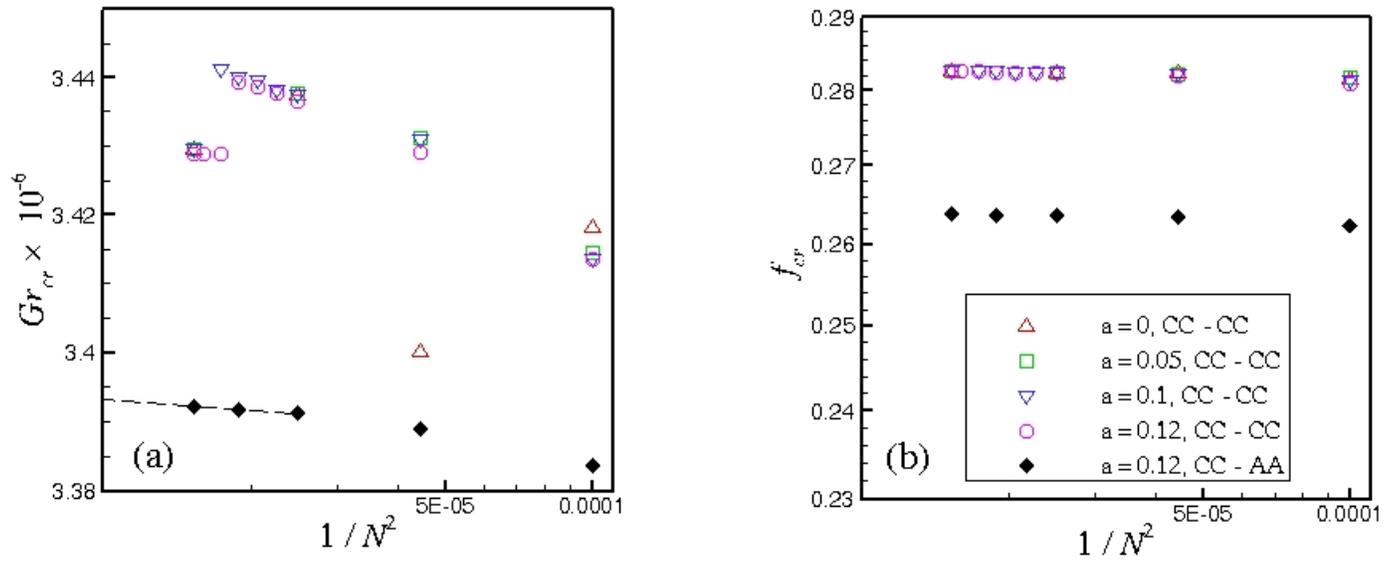

Figure 4. Convergence of the critical Grashof numbers and critical frequencies with the grid refinement. $N$ denotes the number of grid points in one spatial direction.

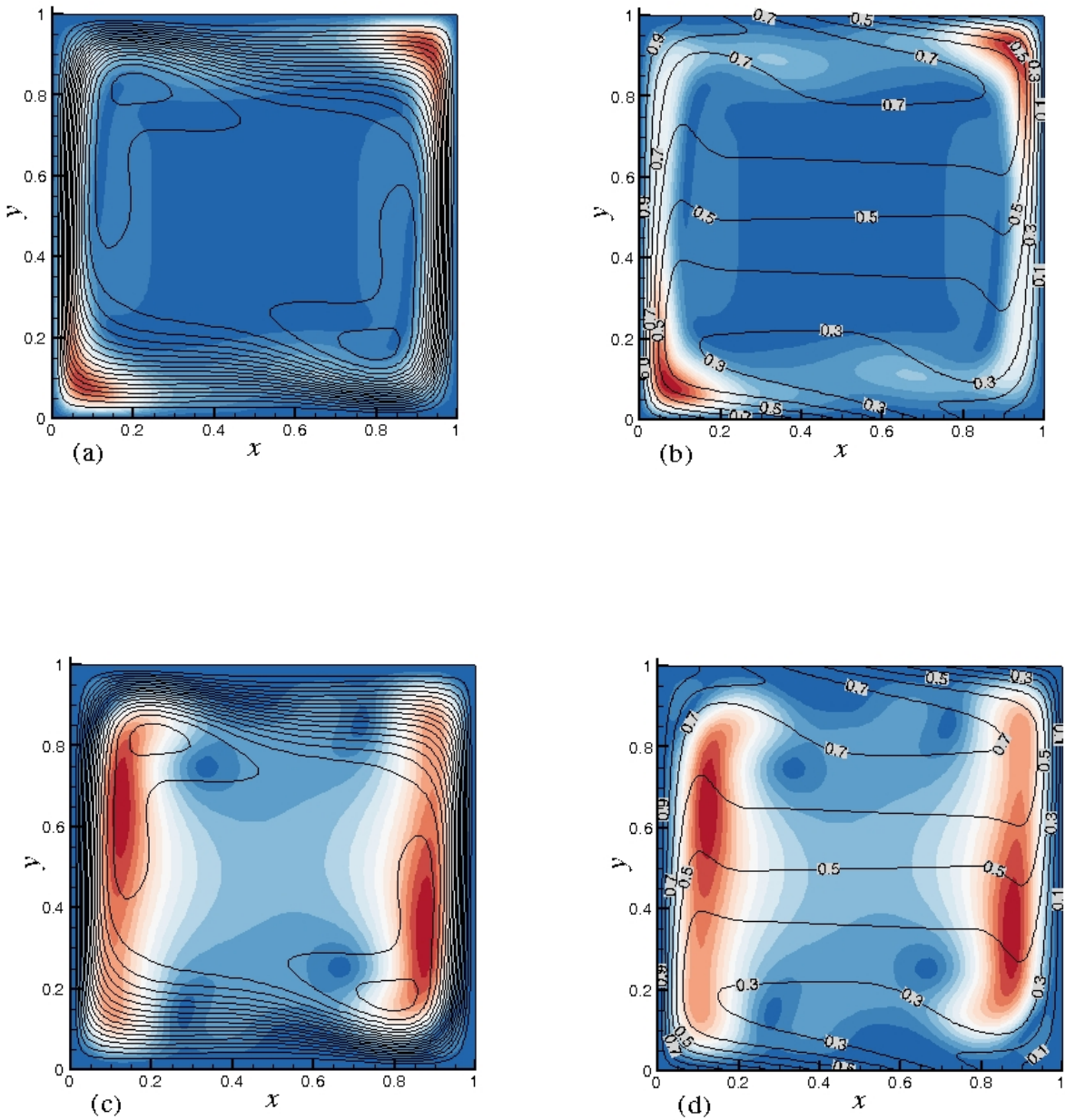

Figure 5. Color plots: amplitude of the temperature (a), (b) and stream function (c), (d) perturbations. Black lines: streamlines (a), (c) and isotherms (b), (d) of the base flow at the critical Grashof number. $Pr = 0.71, Gr_{cr} = 2.9 \times 10^6$. Minimal and maximal values of the stream function in the frames (a) and (c) are $-36.86$ and $0$.

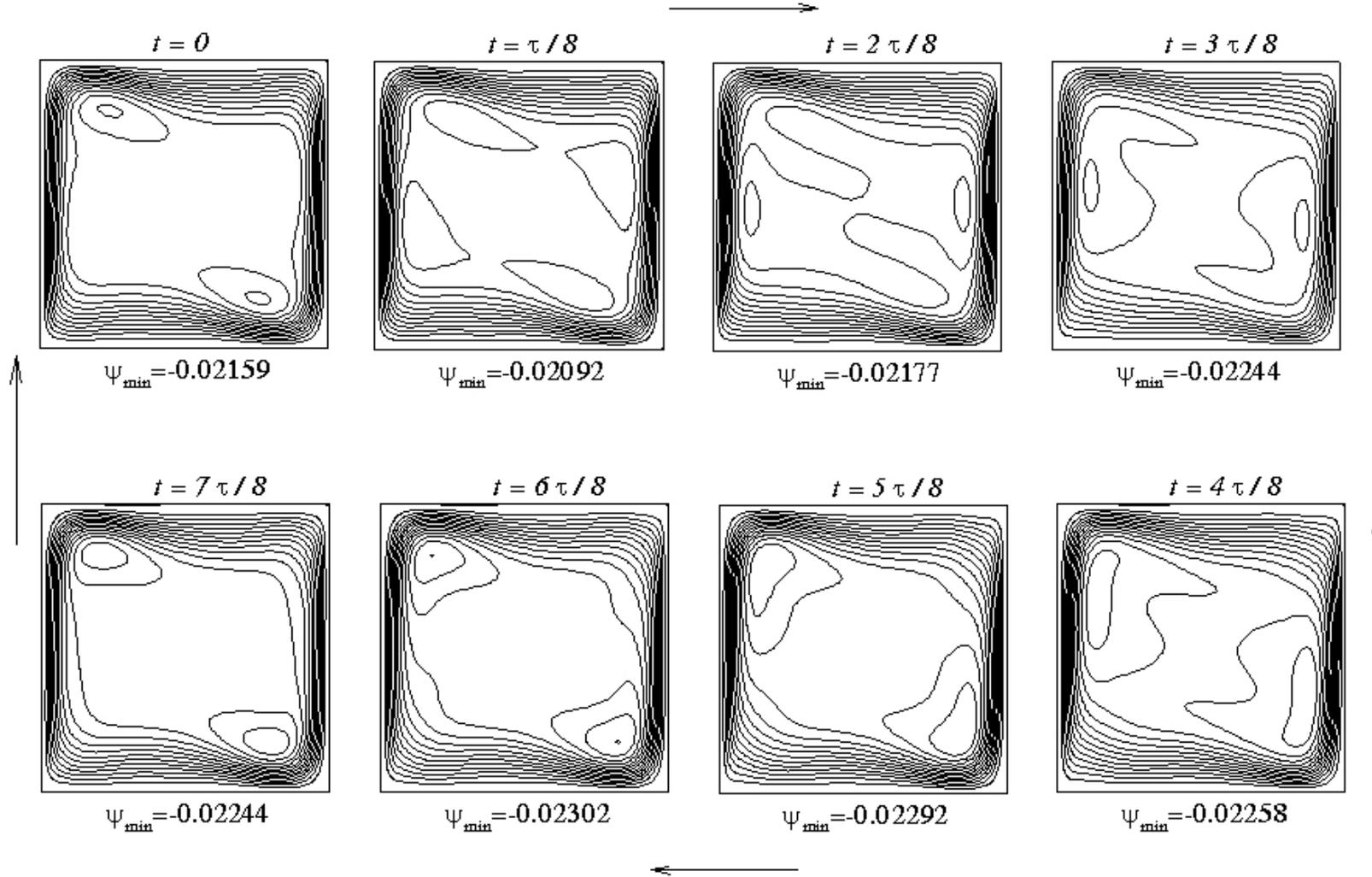

Figure 6. Snapshots of the streamlines of slightly supercritical oscillatory 2D flow obtained as a superposition of the base state with the leading disturbance. The streamlines are equally spaced between the values $-0.023$ and $0$. Axes like in Fig. 4. Animation 1.

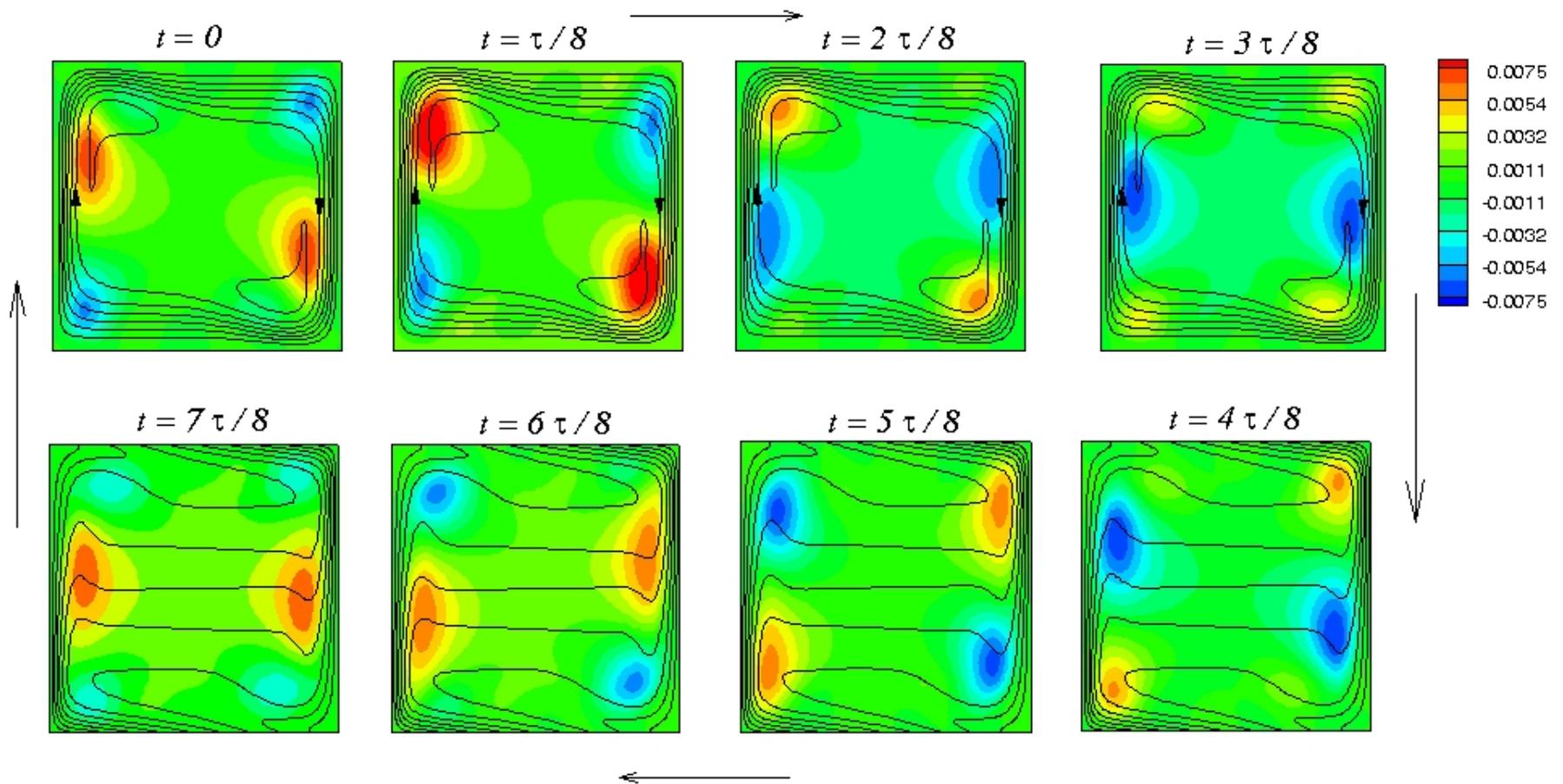

Figure 7. Snapshots of 2D dominant perturbation of the stream function shown on the streamlines (upper row) and isotherms (lower row) of the base 2D state at the critical point. Stream function is equally spaced between the values −35 and 0, and the temperature between 0 and 1. Axes like in Fig. 4. Animation 2.

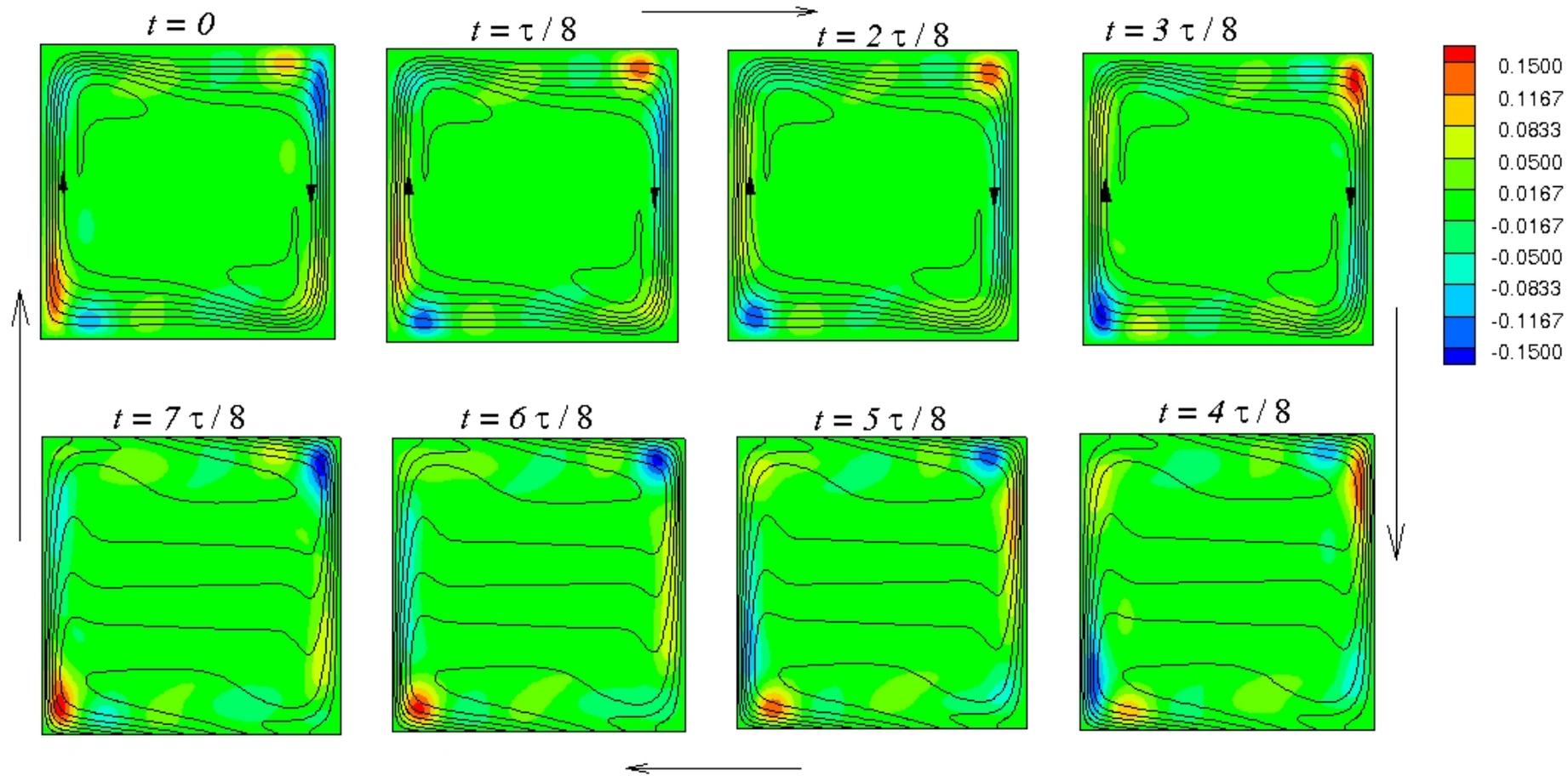

Figure 8. Snapshots of 2D dominant perturbation of the temperature shown on the streamlines (upper row) and isotherms (lower row) of the base 2D state at the critical point. Stream function is equally spaced between the values −35 and 0, and the temperature between 0 and 1. Axes like in Fig. 4. Animation 3.

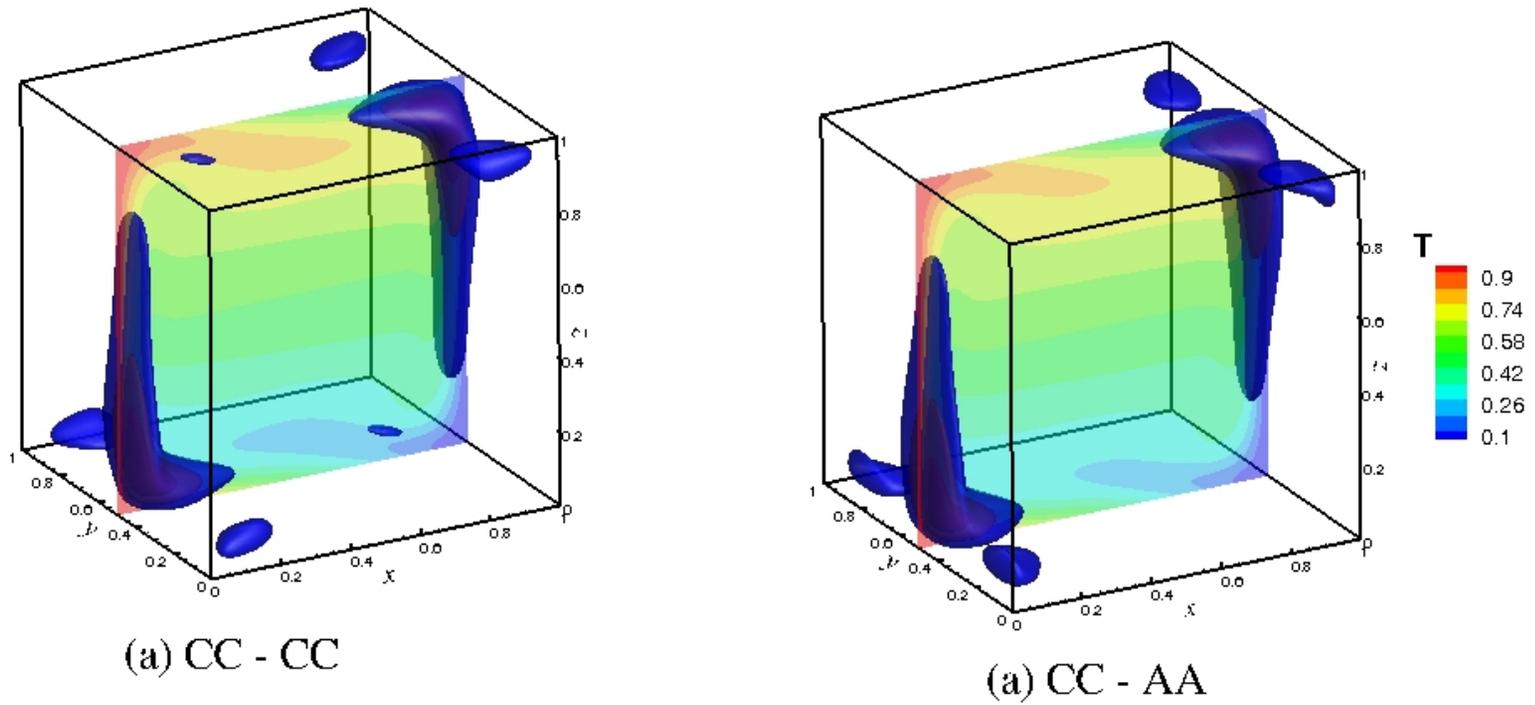

(a) CC - CC                                   (a) CC - AA

Figure 9. Amplitude of the temperature perturbations with isotherms of the base flow at the critical Grashof number. $Pr = 0.71, Gr_{cr} = 3.4 \times 10^6$. (a) CC – CC case, (b) CC – AA case .

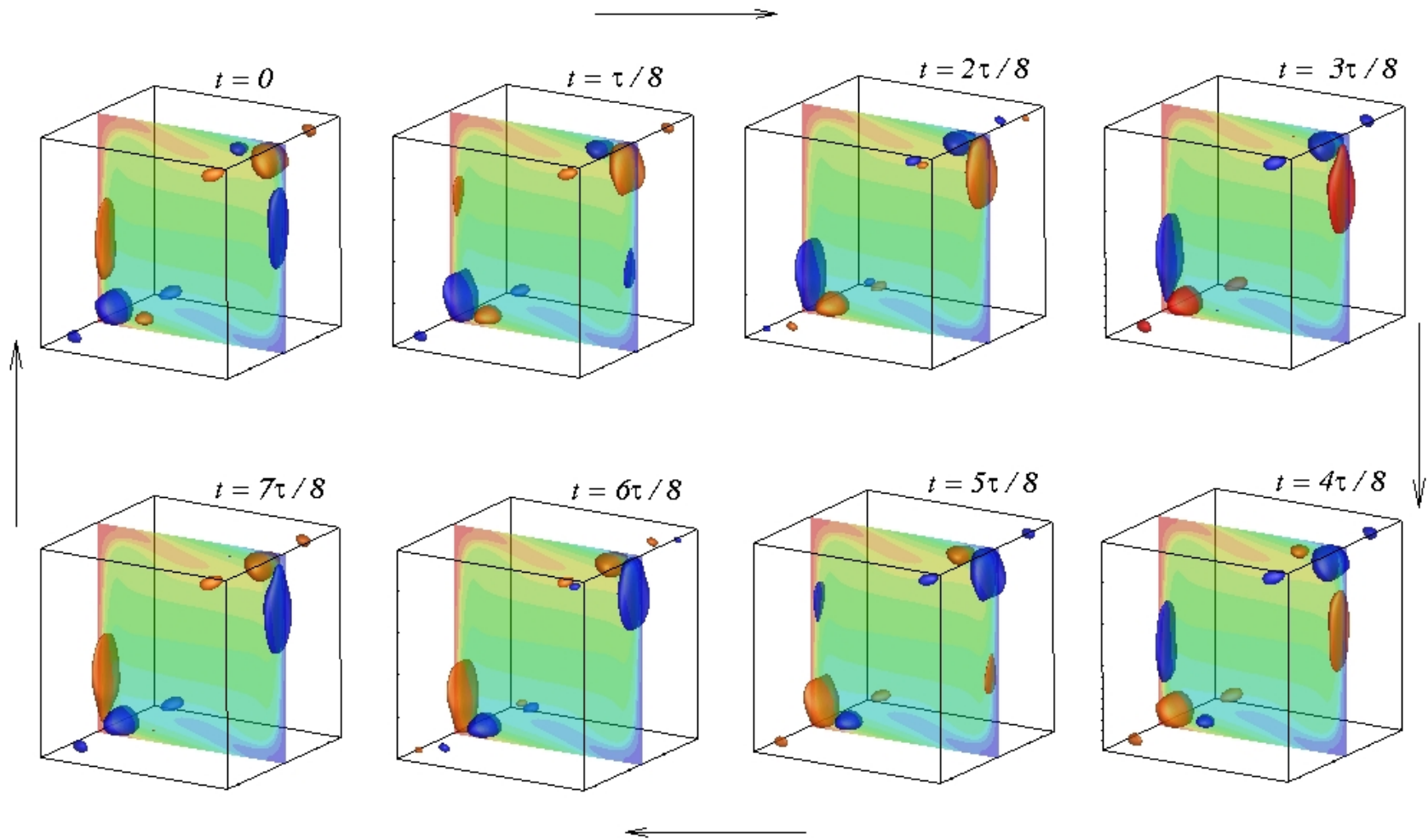

Figure 10. Snapshots of 3D perturbation of the temperature with isotherms of the base flow in the midplane. CC – CC case. Axes like in Fig. 8. Animation 4.

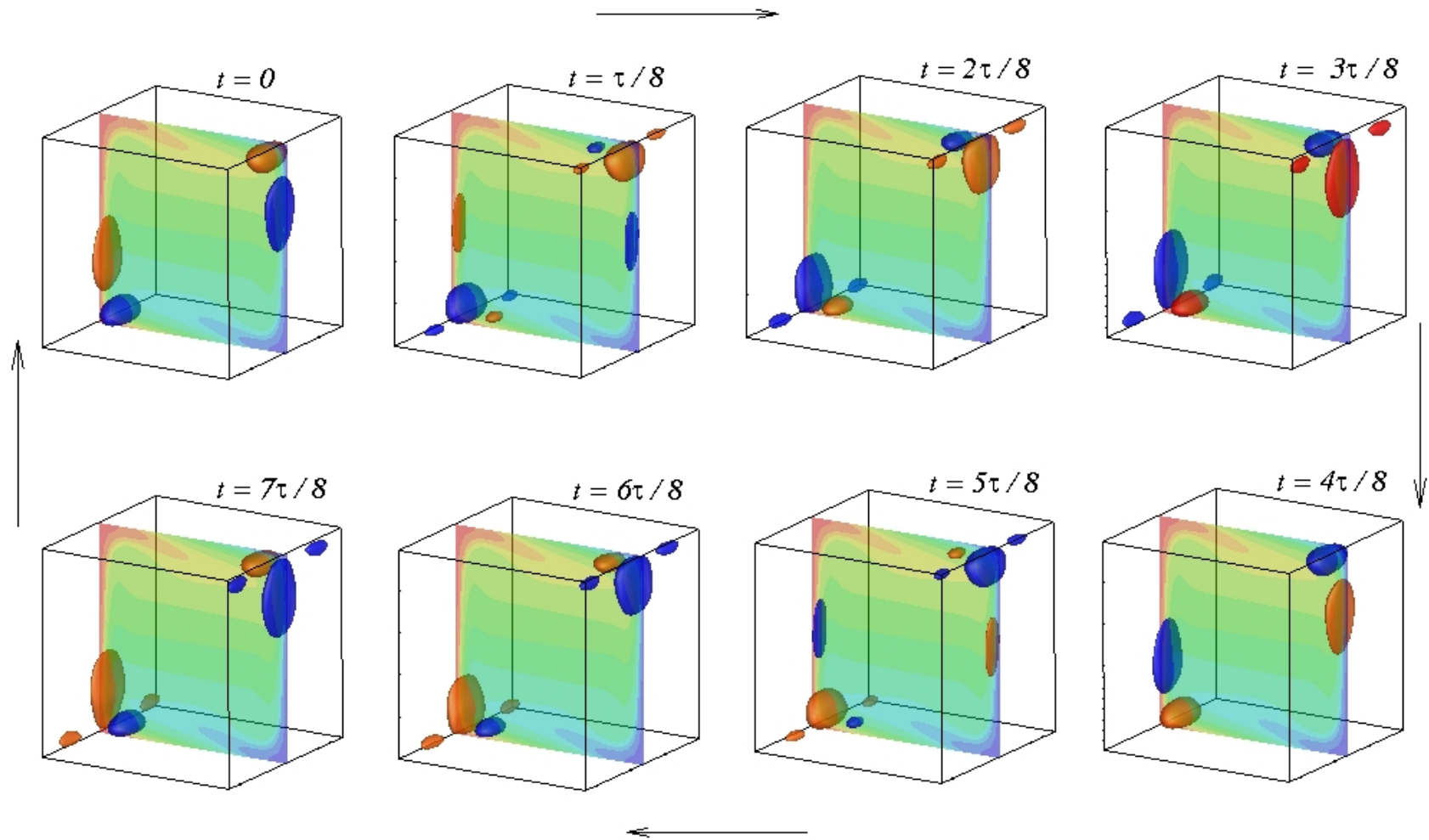

Figure 11. Snapshots of 3D perturbation of the temperature with isotherms of the base flow in the midplane. CC – AA case. Axes like in Fig. 8. Animation 5.

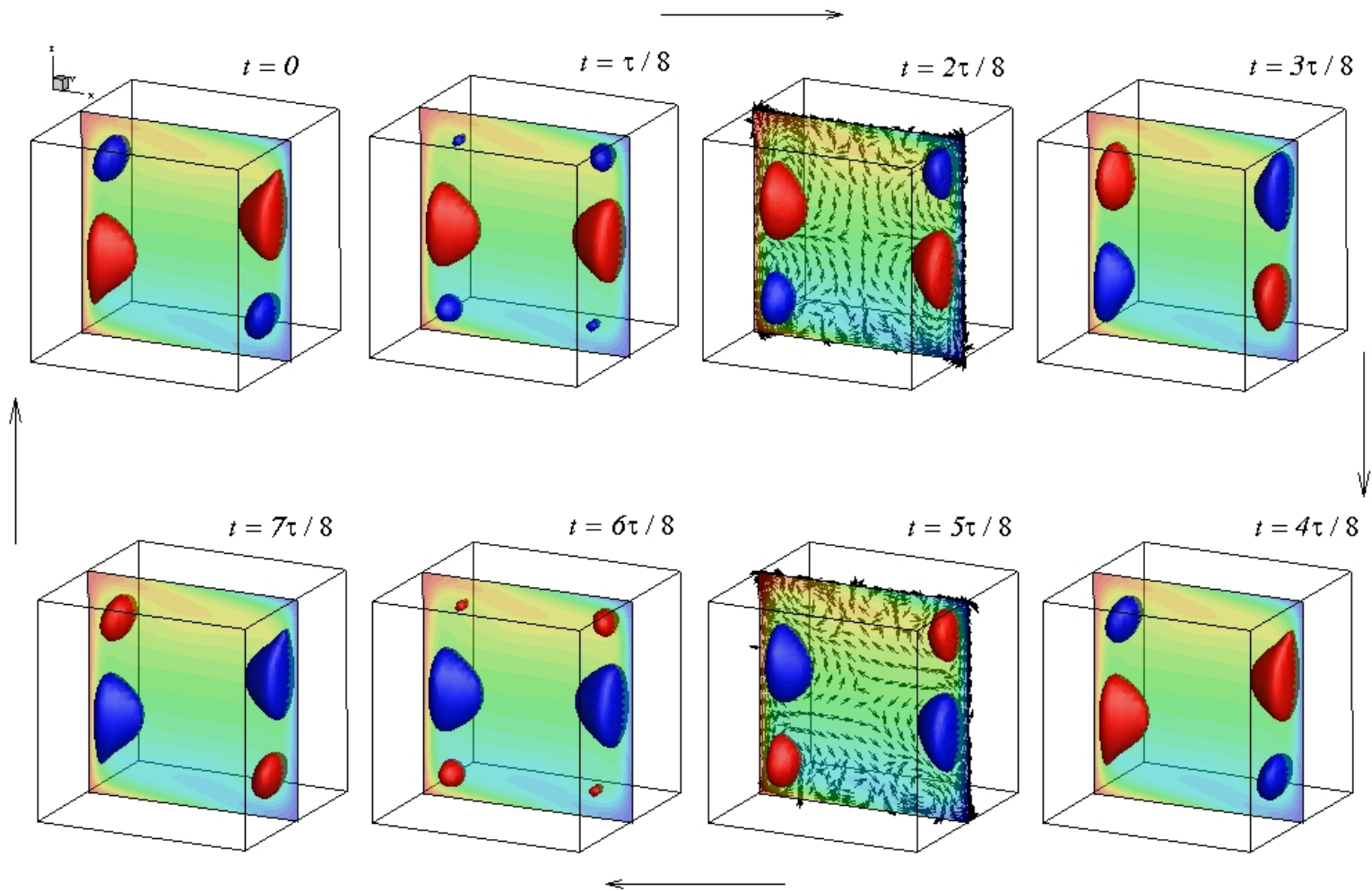

Figure 12. Snapshots of 3D perturbation of the vector potential $\Psi_y$ with isotherms of the base flow in the midplane. $max|\Psi_z| = 2.4 \cdot 10^{-4}$. Levels shown are $\pm 10^{-4}$. Vector plots indicate directions of perturbed part of the flow only. CC − CC case.

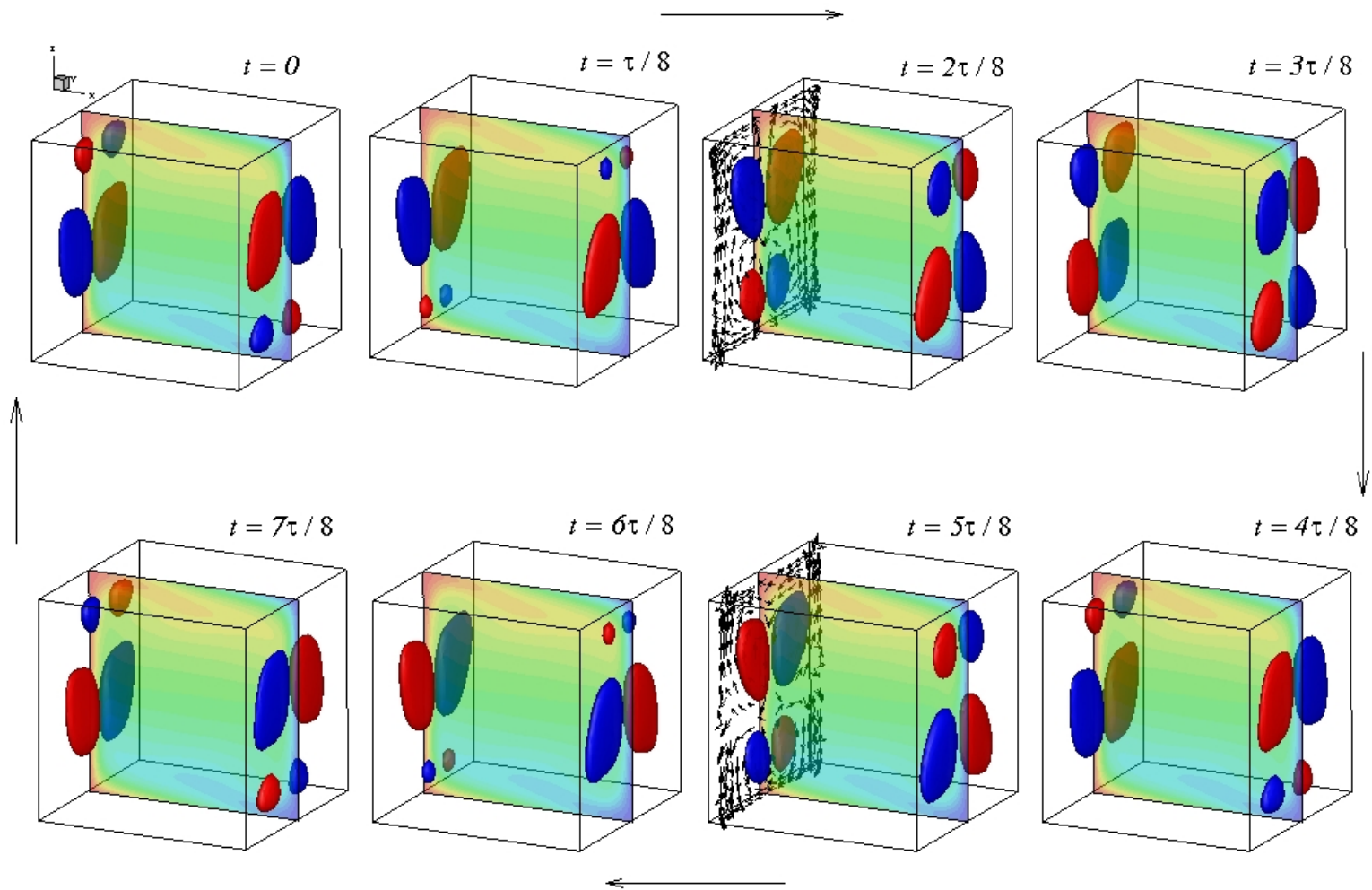

Figure 13. Snapshots of 3D perturbation of the vector potential $\Psi_x$ with isotherms of the base flow in the midplane. $max|\Psi_x| = 2.5 \cdot 10^{-4}$. Levels shown are $\pm 10^{-4}$. Vector plots indicate directions of perturbed part of the flow only. CC – CC case.

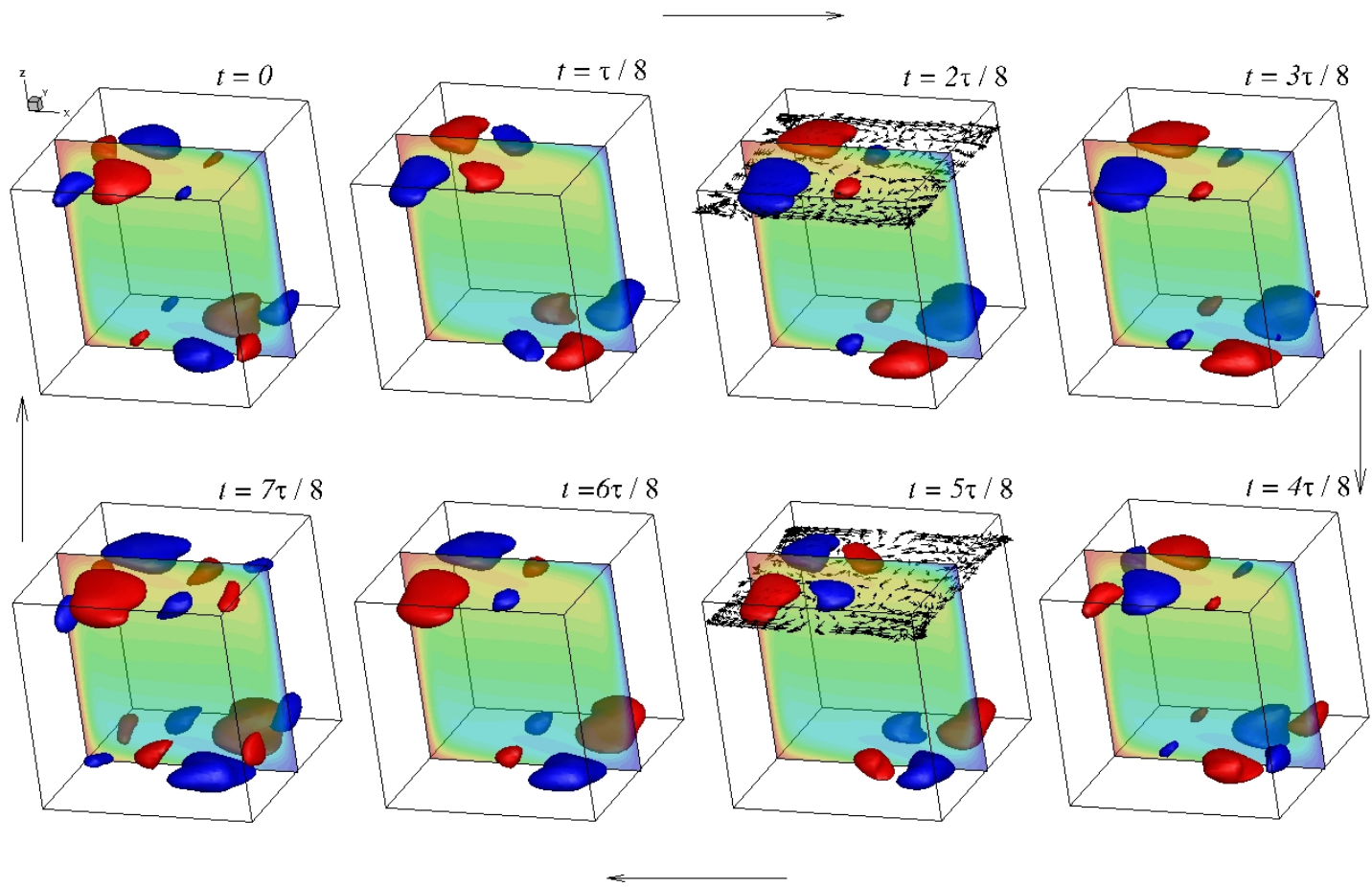

Figure 14. Snapshots of 3D perturbation of the vector potential $\Psi_z$ with isotherms of the base flow in the midplane. Levels shown are $\pm 2 \cdot 10^{-5}$. Vector plots indicate directions of perturbed part of the flow only. CC – CC case. $max|\Psi_z| = 7.5 \cdot 10^{-5}$. Animations 6 and 7.